\documentclass[trackchanges,twocolumn]{aastex63}

\usepackage{CJK}
\usepackage{amsmath}
\usepackage{physics}
\usepackage{fixmath}
\usepackage{upgreek}
\usepackage{textcomp}
\usepackage{amssymb}
\usepackage{tikz}
\usetikzlibrary{arrows}
 \usepackage{booktabs}
 \usepackage{multirow}
\usepackage{tabularx}
\usepackage{booktabs,multirow}

\definecolor{DarkGreen}{rgb}{0.0,0.4,0.0}  

\usepackage[normalem]{ulem}
\usepackage{soul}
\usepackage{cancel}

\submitjournal{ApJL}

\shorttitle{rotation of magnetic flux rope}
\shortauthors{Zhou et al.}

\received{\today}
\revised{\today}
\submitjournal{APJL}

\begin{document}
\begin{CJK*}{UTF8}{gbsn}
\title{The Rotation of Magnetic Flux Rope Formed during Solar Eruption}
\correspondingauthor{Zhenjun Zhou, Chaowei Jiang}
\email{zhouzhj7@mail.sysu.edu.cn, chaowei@hit.edu.cn}

\author[0000-0001-7276-3208]{Zhenjun Zhou(周振军)}
\affiliation{Planetary Environmental and Astrobiological Research Laboratory (PEARL), School of Atmospheric Sciences, Sun Yat-sen University, Zhuhai, China}
\affiliation{CAS Key Laboratory of Geospace Environment, University of Science and Technology of China, Hefei, Anhui 230026, China}
\affiliation{CAS Center for Excellence in Comparative Planetology, China}
\author[0000-0002-7018-6862]{Chaowei Jiang}
\affiliation{Institute of Space Science and Applied Technology, Harbin Institute of Technology, Shenzhen 518055, People's Republic of China}
\author[0000-0003-4618-4979]{Rui Liu}
\affiliation{CAS Key Laboratory of Geospace Environment, University of Science and Technology of China, Hefei, Anhui 230026, China}
\author[0000-0002-8887-3919]{Yuming Wang}
\affiliation{CAS Key Laboratory of Geospace Environment, University of Science and Technology of China, Hefei, Anhui 230026, China}
\author[0000-0001-6804-848X]{Lijuan Liu}
\affiliation{Planetary Environmental and Astrobiological Research Laboratory (PEARL), School of Atmospheric Sciences, Sun Yat-sen University, Zhuhai, China}
\author[0000-0002-4721-8184]{Jun Cui}
\affiliation{Planetary Environmental and Astrobiological Research Laboratory (PEARL), School of Atmospheric Sciences, Sun Yat-sen University, Zhuhai, China}

\begin{abstract}
The eruptions of solar filaments often show rotational motion about their rising direction, but it remains elusive what mechanism governs such rotation and how the rotation is related to the initial morphology of the pre-eruptive filament (and co-spatial sigmoid), filament chirality, and magnetic helicity. The conventional view regarding the rotation as a result of a magnetic flux rope (MFR) undergoing the ideal kink instability still has confusion in explaining these relationships. Here we proposed an alternative explanation for the rotation during eruptions, by analyzing a magnetohydrodynamic simulation in which magnetic reconnection initiates an eruption from a sheared arcade configuration and an MFR is formed during eruption through the reconnection. The simulation reproduces a reverse S-shaped MFR with dextral chirality, and the axis of this MFR rotates counterclockwise while rising, which compares favorably with a typical filament eruption observed from dual viewing angles. By calculating the twist and writhe numbers of the modeled MFR during its eruption, we found that accompanied with the rotation, the nonlocal writhe of the MFR's axis decreases while the twist of its surrounding field lines increases, and this is distinct from the kink instability, which converts magnetic twist into writhe of the MFR axis. 
\end{abstract}

\emph{Online-only material}: animations, color figures
\section{Introduction}

Solar eruptions often show rotational motion, especially in filament eruptions, which are commonly believed to be the manifestation of erupting magnetic flux ropes (MFRs). For example, during some filament eruptions, the observation of a bunch of helical threads that appear to wind around a central axis with rotation motion is reminiscent of an MFR. The rotation motion plays an important role in reconfiguring the erupting magnetic field. For instance, on the one hand, it may yield a possible magnetic reconnection of the erupting MFR with the surrounding field ruining the coherence of the erupting MFR, even resulting in a failed eruption \citep{Zhou2019}. On the other hand, it can persistently modulate the axis direction of the subsequent coronal mass ejection (CME), and change the southward component of the interplanetary magnetic field, which therefore makes the prediction of potential geoeffectiveness more challenging \citep{Yurchyshyn2009}.

The twisting and rotating features indicate that their underlying magnetic fields carry currents and possess magnetic helicity, and the direction of rotation is closely related to the sign of helicity. Indeed, observations of filament eruptions show that there is a one-to-one correlation between the rotation direction and the filament chirality (or sign of helicity of the corresponding magnetic field); sinistral (dextral) filaments with positive (negative) helicity rotate clockwise (counterclockwise) when viewed from above \citep{Green2007}. Furthermore, the pre-eruptive morphology of filaments corroborates an moderate hemispheric preference, namely, filaments of forward (reverse) S shape are usually located in the southern (northern) hemisphere and have a positive (negative) helicity \citep{Rust1996,Zhou2020}. Besides, there are coronal loops presenting S shape known as sigmoids observed in extreme ultraviolet (EUV) and soft X-ray (SXR) passbands \citep{Cheng2017}, and in many events the sigmoid is found to be co-spatial roughly with the pre-eruptive filament. 

To explain the relationships between the rotation direction of the MFR as it erupts, the field chirality, and the associated filament (sigmoid) morphology, 
\citet{Green2007} invoked the theory based on the Titov and D{\'e}moulin (T\&D) model \citep{Titov1999} which assumes an arched MFR pre-existing before eruption, 
and argued that the observations agree well with the T\&D model. In the T\&D model, the observed sigmoid is considered to be a thin current layer formed in the bald patch separatrix layer (BPSS) or hyperbolic flux tube (HFT) in the wake of the rising flux rope, 
and \citet{Green2007} suggested that the observed relationship between the filament chirality and its rotation direction is a manifestation of ideal MHD instability of the MFR, during which the magnetic twist is converted into the writhe of the axis (thus rotation of the MFR's axis). However, \citet{Torok2010} found that for the T\&D's MFRs, the relation between writhe and the projected S shape of MFRs is not unique, since the writhe depends largely on the height of the MFRs and on the presence or absence of dips in the middle of the MFR, rather than the transformation of their twist helicity into writhe helicity as is often assumed.

In this Letter, we proposed an alternative explanation for the eruption rotation, which is more uniformly consistent with the observations, by using a reconnection-initiated eruption model in which an MFR does not need to exist before eruption but is formed during eruption through reconnection within a sheared arcade configuration. Our explanation is developed based on a recent fully 3D magnetohydrodynamic (MHD) simulation \citep{Jiang2021} which demonstrates for the first time that the runaway tether-cutting reconnection alone can initiate solar eruption within a single arcade as sheared by photospheric motion. In the simulation, the MFR is formed during the eruption through reconnection of the sheared arcade. Here we will analyze the morphology, chirality, and rotation direction of the erupting MFR in the simulation, and compare them with the observations of a typical filament eruption. Our results show that at the onset of the eruption, the MFR is built up with a reverse S shape, and the top of the MFR shows a significant counterclockwise rotation immediately after the initiation of the eruption, which are entirely consistent with the observations. Furtherly, by a quantitative measurement of writhe and twist of the MFR in the simulation, we found that there a transfer of writhe to twist in the MFR during eruption, which is distinct from the previous theory based on kink instability. 

\section{Observation of a typical filament eruption} \label{obs}

\begin{figure*} 
      \vspace{-0.03\textwidth}    
      \centerline{\hspace*{0.00\textwidth}
      \includegraphics[width=1.0\textwidth,clip=]{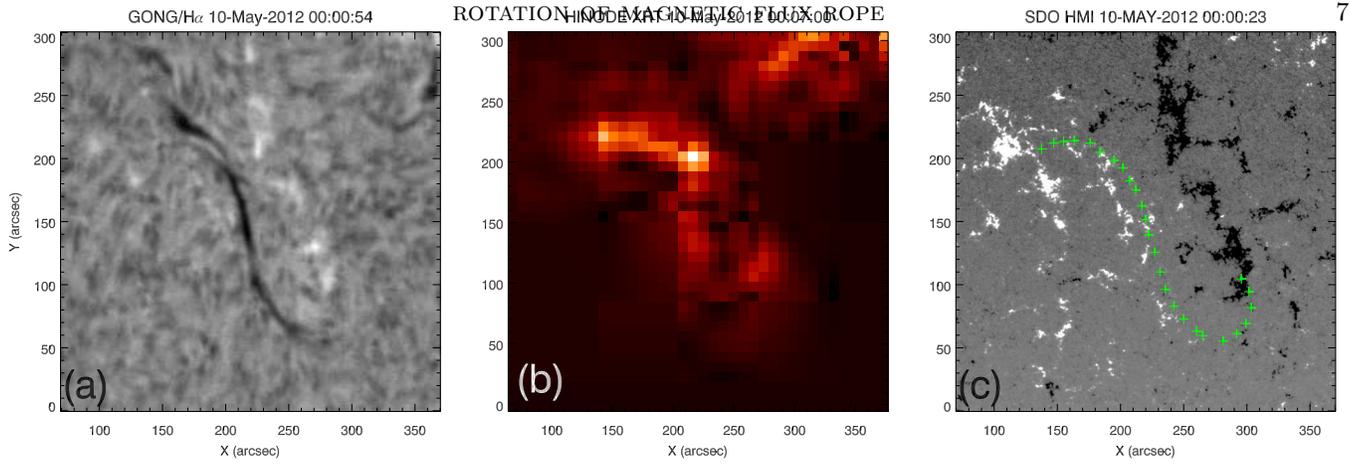}
      }
\caption{Panel (a) shows a reverse S-shaped filament observed by GONG/H\(\alpha \) on 2012 May 10 at 00:00 UT. The corresponding sigmoid observed by Hinode/XRT is shown in (b). In Panel (c), the green plus symbols outlining the EUV filament spine are projected onto an HMI magnetogram of local $B_r$.} \label{f1}
\end{figure*}

We first take a typical filament eruption, which occured in NOAA active region (AR) 11475 on May 10th, 2012, as an example to illustrate the relationship between the orientation of the S-shaped morphology of the filament (and its associated sigmoid), the filament chirality, and the rotation direction during its eruption. The filament is well observed by $\mathrm{H}\upalpha$ image in the 6563 {\AA} wavelength from Global Oscillation Network Group \citep[GONG;][]{Hill1994}, showing an inverse S shape located on the solar disk center (S15W15) as seen from the Earth view (Figure~\ref{f1}(a)). It has a wider appearance in the He II 304 {\AA} from dual-perspective imaging observations from Solar Dynamics Observatory (SDO)/Atmospheric Imaging Assembly \citep[AIA;][]{Lemen2012} and Solar TErrestrial RElations Observatory (STEREO)-A/ Extreme UltraViolet Imager \citep[EUVI;][]{Wuelser2004}. A sigmoid co-spatial with the filament is observed with the X-ray telescope \citep[XRT;][]{Golub2007} onboard Hinode (Figure~\ref{f1}(b)). 

The chirality of the filament can be determined with the help of the radial magnetogram provided by the Heliospheric and Magnetic Imager \citep[HMI;][]{Schou2012} onboard SDO. Figure~\ref{f1}(c) shows that overall this AR has a bipolar configuration where the two opposite polarities are aligned northeast$-$southwest, and the filament observed in the SDO/AIA 304 {\AA} (Figure~\ref{f2}(d)) is outlined by the green plus symbols as overplotted on the magnetogram. 
The footpoints of the magnetic field supporting the filament can be located by where the filament plasma flows down to the solar surface. The observed right-skewed drainage sites relative to the PIL implies that the filament chirality is dextral \citep{Chen2014, Zhou2020}. 
Meanwhile, during its eruption, the apex part of the filament displays CCW rotation, which can be seen in both SDO and STEREO-A observations (Figure~\ref{f2}(b),(d), and accompanied animation). This event is a well-observed example supporting a strong one-to-one relationship during the eruption as found by \citet{Zhou2020}; sinistral/dextral filaments rotate clockwise (CW)/counterclockwise (CCW) when viewed from above, and the morphology of the filament and related sigmoid both exhibit a forward (reverse) S shape.

\section{MHD Simulation of an erupting MFR} \label{Mod}
\citet{Jiang2021} performed a high accuracy, fully 3D MHD simulation and established a fundamental mechanism behind solar eruption initiation: a bipolar field driven by quasi-static shearing motion at the photosphere can form an internal current sheet, followed by fast magnetic reconnection that triggers and drives the eruption. In this mechanism, an MFR is built up during the eruption, and here we focus on the evolution of the erupting MFR by using the simulation run like the one in \citet{Jiang2021}, but with a lower resolution than the original ones. The simulation solves the full set of MHD equations with both solar gravity and plasma pressure included, and starts from a bipolar potential magnetic field and a hydrostatic plasma stratified by solar gravity with typical coronal temperature. Then shearing flows along the PIL, which are implemented by rotating the two magnetic polarities at the photosphere in the same CCW direction, are applied on the bottom boundary to energize the coronal field until an eruption is triggered, and after then the surface flow is stopped. During the quasi-static evolution phase as driven by the shearing motion, a current sheet is gradually built up. Since no explicit resistivity is used in the MHD model, magnetic reconnection is triggered when the current sheet is sufficiently thin such that its width is close to the grid resolution, owing to the implicit, grid-dependent numerical resistivity. For more details of the simulation settings, the readers are referred to \citet{Jiang2021}. In that paper, the simulation is managed to be of very high resolutions with Lundquist number achieving $\sim 10^5$ for a length unit (approximately 10 Mm). Therefore, the plasmoid instability is triggered in the current sheet and the magnetic topology becomes extremely complicated in small scales along with formation of the large-scale MFR. Such a complexity substantially complicates our analysis of the large-scale evolution associated with the erupting MFR, thus in this paper we used a lower-resolution run (corresponding to a Lundquist number of $\sim 10^3$). In the lower-resolution run, the amount of shearing time before the eruption onset is a little less than that needed in the high-resolution run, because the current sheet required for triggering reconnection is thicker and thus needs less shear (which has been shown in \citet{Jiang2021} with four different resolutions). That said, the basic evolution of the MFR during the eruption is not changed as compared to the high-resolution run, except that the small-scale complex structure will not arise. Moreover, with the lower resolution, we can run the simulation longer and thus follow a longer evolution of MFR.

\section{Comparison of Simulation and Observation} \label{sec:com}
\begin{figure*} 
      \vspace{-0.03\textwidth}    
      \centerline{\hspace*{0.00\textwidth}
      \includegraphics[width=1.0\textwidth,clip=]{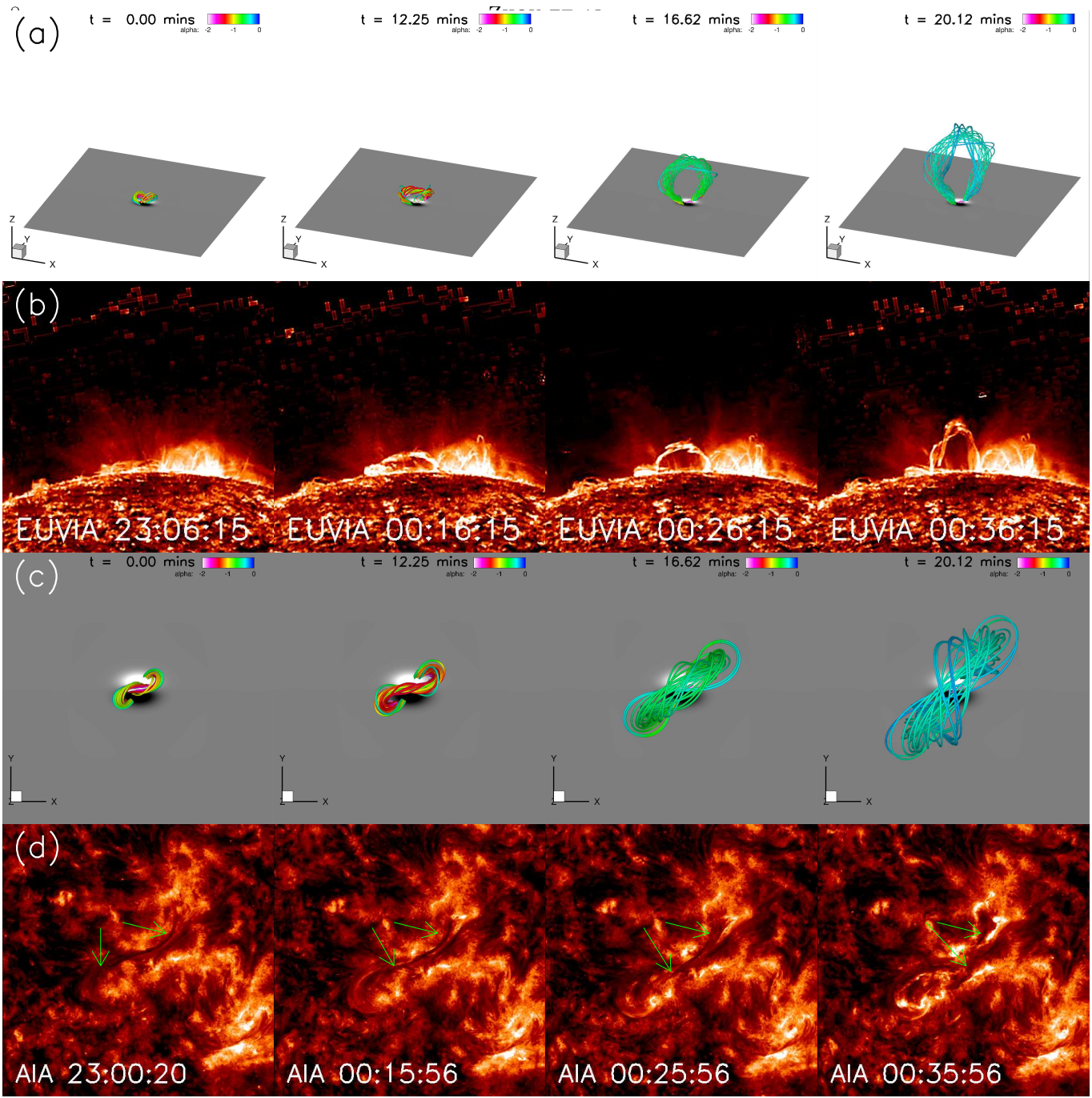}
      }
\caption{ Comparison of the filament eruption on 10 May 2012 with the numerical simulation of \citet{Jiang2021}. Panels (a) and (c) Side and top views of the MFR eruption sequentially in the simulation. The colored thick lines represent the MFR's core magnetic field lines, and the colors denote the value of the nonlinear force-free factor defined as $\alpha$=$\bf J \cdot \bf B$ /$B^{2}$,
where $\bf J$ is the current density and $\bf B$ is the magnetic field. 
Panels (b) and (d) provide sequential snapshots of the filament eruption from the limb view in STEREO-A/EUVI 304 {\AA} and disk view in SDO/AIA 304 {\AA}, respectively, showing a CCW rotation of the filament, all the images here from STEREO and SDO has been rotated to accommodate with the simulation for comparison purpose. The video duration is 24s.
(An animation of this figure is available.)} \label{f2}
\end{figure*}

Figure~\ref{f2} compares the process of the filament eruption from 23:00 UT on May 9 to 00:36 UT on May 10 observed by STEREO-A/EUVI, and SDO/AIA with magnetic field evolution seen in two different views from the MHD simulation. At the onset of the eruption (Note that here for the simulation, $t=0$ is reset to be the onset time of the simulated eruption), the core magnetic field of the newly formed MFR, which is built up through reconnection of the two sets of sheared arcades, presents a continuous reverse S-shaped sigmoid from the top view, and subsequently exhibits a significant CCW rotation during the eruption (Figure~\ref{f2}(c)). From the side (limb) view, the low-lying flux rope rises up quickly, yielding a nearly circular shape, and further with the CCW rotation the shape is transformed into an oval (Figure~\ref{f2}(a)). Therefore, the evolving morphology of the erupting flux rope in the simulation agrees well with that of the erupting filament in the dual-perspective observations from STEREO-A and SDO.

\begin{figure*} 
     \vspace{-0.0\textwidth}    
     \centerline{\hspace*{0.00\textwidth}
     \includegraphics[width=1.0\textwidth,clip=]{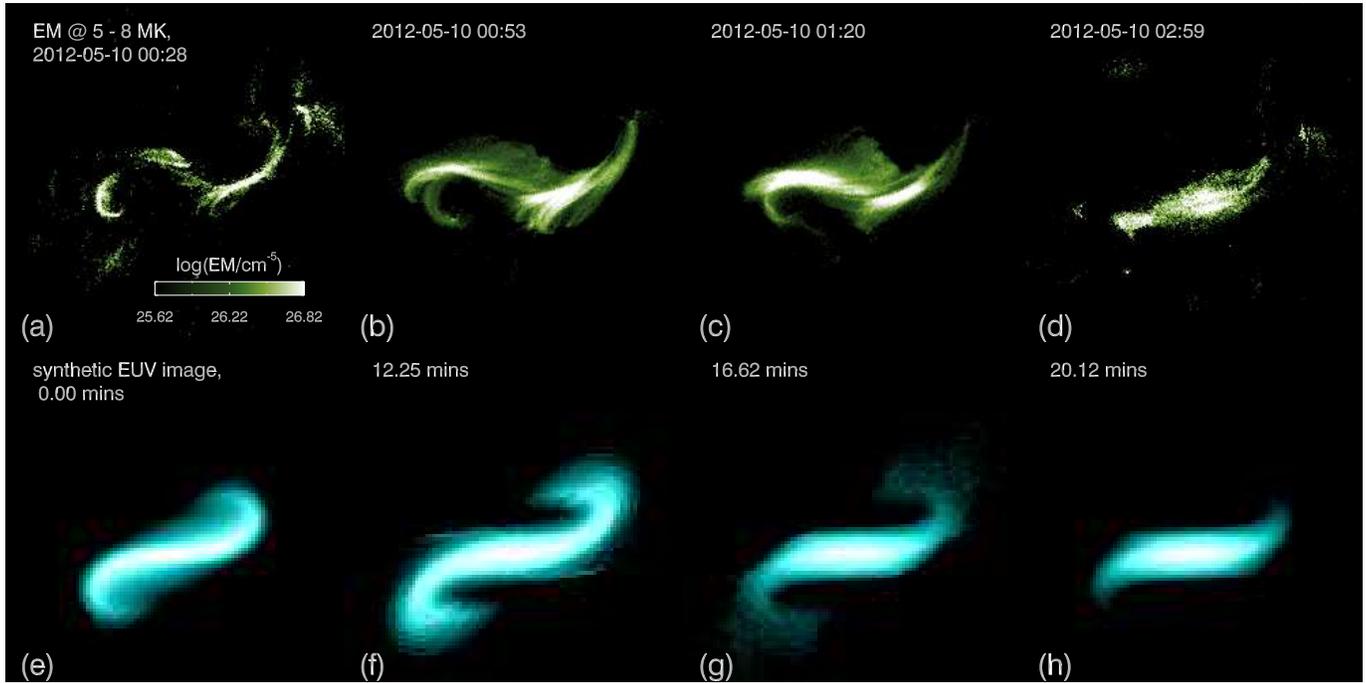}
               }
\caption{Thermal evolution of the solar eruption and simulation. EM maps ranging from 5 - 8 MK (a)-(d) for the filament eruption and synthetic EUV images (e)-(h) for the numerical simulation. The video duration is 18s.
(An animation of this figure is available.)} \label{f3}
\end{figure*}

To compare the thermal morphology of the observed sigmoid with that of the simulation,
we deduce the thermal evolution of this eruption based on imaging data from six AIA EUV passbands, including 131{\AA} (Fe XXI, $\sim$11 MK; Fe VIII, $\sim$0.4 MK), 94{\AA} (Fe XVIII, $\sim$7.1 MK; Fe X, $\sim$1.1 MK), 335{\AA} (Fe XVI, $\sim$2.5 MK), 211{\AA} (Fe XIV, $\sim$2.0 MK), 193{\AA} (Fe XII, $\sim$1.6 MK; Fe XXIV,  $\sim$17.8 MK), and 171{\AA} (Fe IX, $\sim$0.6 MK) \citep{ODwyer2010}. 
We use a sparse inversion code \citep{Cheung2015,Su2018} to calculate the emission measure (EM) as a function of temperature from  AIA imaging data.
EM is the integral of electron density squared over the emitting volume, it gives the amount of plasma emission at a given temperature.
Due to the limitation of HINODE/XRT observations, evolution of sigmoid can't be followed up in SXR passbands, we use this EM map as a substitution.
The EM maps over the temperature range from 5 -8 MK show a clear sigmoid-to-arcade transformation during the eruption (Figure~\ref{f3}(a)-(d)). Initially, a coaxial, bright feature appears in the wake of this rising filament, broadens as a sigmoidal shape, and finally, evolves into an arcade shape.
This sigmoidal emission pattern is expected to be due to the heating in the current-carrying magnetic fields \citep{Kliem2004,Gibson2006}.
To visualize current-carrying field lines in the simulation for comparison, a method for the synthesis of mock coronal images is utilized. It calculates line-of-sight integrals of the proxy emissivity 
from the value of $j^2$ (square of the current density) averaged along magnetic field lines  \citep{Cheung2012}.
From the synthetic images (Figure~\ref{f3}(e)-(h) and accompanied movie), an inverse sigmoidal shape forms before the eruption and then it broadens with the expansion of the erupting field. In the end, the two elbows of the sigmoid fade away and become indistinguishable from the ambient as shown in Figure~\ref{f3}(g) and (h). 
Compared to the heating only along current sheets at the interface between the helical core field (e.g., MFR) and the ambient field \citep{Kliem2004,Gibson2006}, 
this result provides an alternative scenario that the sigmoidal emission pattern is due to the heating as line-of-sight integrals of $j^2$ from both the current sheet and the nearby regions with intense currents.

\section{Evolution of twist and writhe} \label{sec:evo}
To understand the variation of the MFR's morphology during its rising, we investigate the evolution of two parameters, namely, the writhe number ($\mathcal{W}_r$) and the twist number ($\mathcal{T}_w$), which characterize quantitatively the helical deformation of the MFR axis and how much the field lines winding about the MFR axis, respectively. Based on the simulation data, these two parameters are computed using the following methods.  \par

\begin{figure*} 
     \vspace{-0.0\textwidth}    
     \centerline{\hspace*{0.00\textwidth}
     \includegraphics[width=1.0\textwidth,clip=]{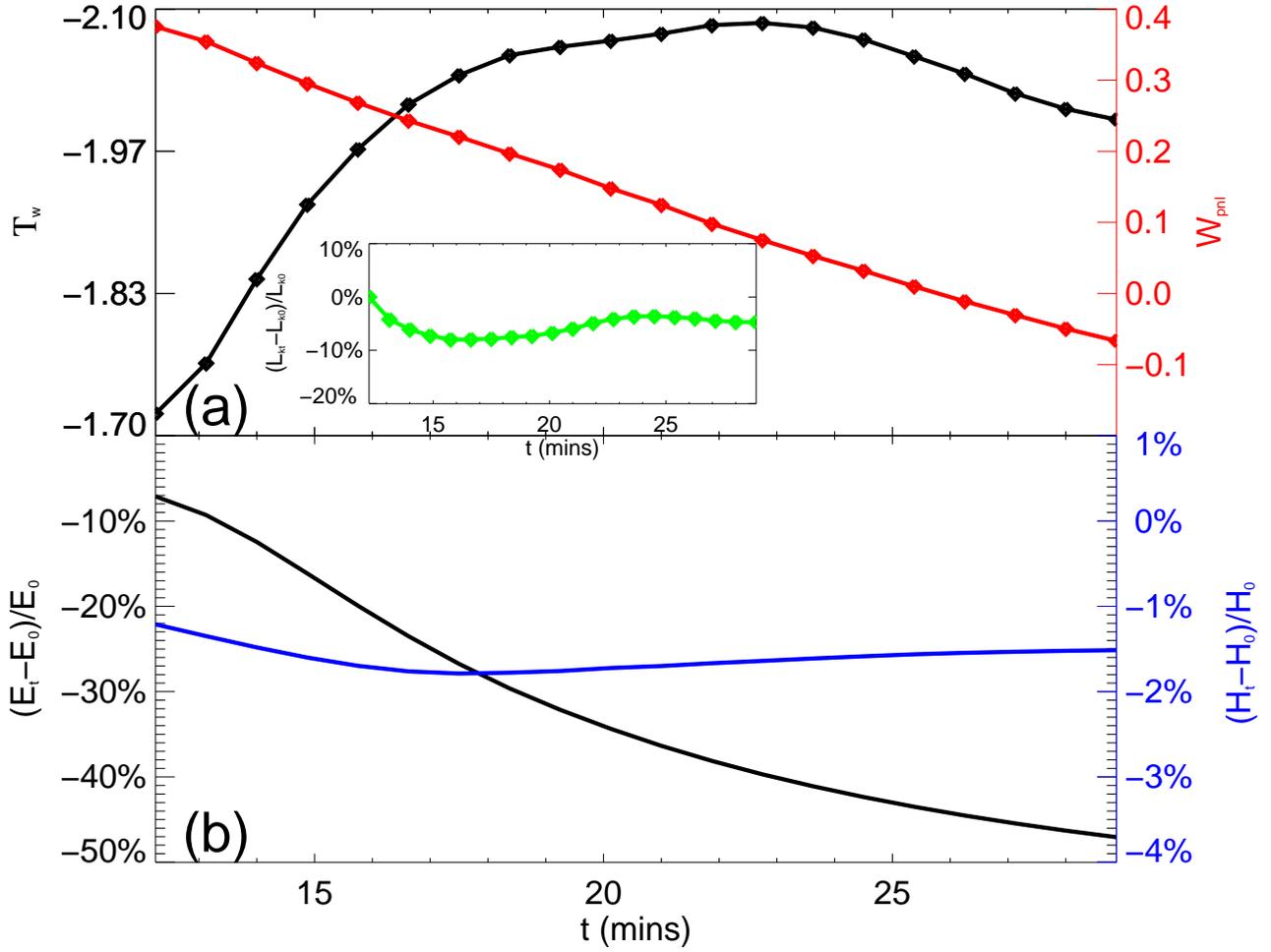}
               }
\caption{Evolution of the parameters in the simulation. Top panel (a): temporal evolution of twist (black line) and nonlocal writhe (red line) of the flux rope during its eruption. In the inset panel, the green line shows the temporal changing rate of linking number ($\mathcal{L}_{k}$); Bottom panel (b): the temporal change ratios of free magnetic energy E (black line) and relative magnetic helicity H (blue line) compared with that of the eruption onset. For reference, at onset of the eruption the magnetic free energy  is $0.91$ of the corresponding potential magnetic energy, and the relative helicity is $-0.06$ as normalized by the square of the total unsigned magnetic flux. } \label{f5}
\end{figure*}

For an open curve like the axis of an MFR with both endpoints on a bottom plane (e.g., the photosphere), its temporal evolution of writhe is difficult to be quantified \citep{Linton1998}. \citet{Berger2006} proposed a modified writhe expression termed the polar writhe to distinguish from the pre-existing closed curve definition. The bottom plane has $\hat{\mathbf{z}}$ as normal. Along the $z$-direction, the open curve is split into several pieces at turning points (extrema in z-direction). The coiled geometric quantity of each individual piece is the local polar writhe ($\mathcal{W}_{pl}$), where the global geometric relations between the pieces is described by the nonlocal polar writhe ($\mathcal{W}_{npl}$). The polar writhe ($\mathcal{W}_p$) is the sum of local and non-local components. This nonlocal component is useful in interpreting the presence of S shape in the corona \citep{Torok2010}. It can be calculated as \citep{Berger2006}:

\begin{equation} 
\mathcal{W}_{pnl}(\textbf{r})=\mathop{\sum_{i=1}^{n+1}\sum_{j=1}^{n+1}}_{i\neq j} \frac{\sigma_{i}\sigma_{j}}{2\pi}\int_{z^{min}_{ij}}^{z^{max}_{ij}}\frac{d\Theta_{ij}}{dz}dz, \label{eqwr}
\end{equation} 

where i,j are two different pieces, $\sigma_{i} =+1$ if this piece is moving upward, $\sigma_{i} =-1$ if this piece is moving downward, and $n$ is the number of the turning points. In the integration,
let the relative position vector at height z be $\mathbf{r}_{ij}(z) = x_j (z) - x_i(z)$. Note that $\mathbf{r}_{ij}(z)$ is parallel to the xy plane.
$\Theta_{ij}$ is the orientation of this vector with respect to the x axis, $z^{min}_{ij}$ and $z^{max}_{ij}$ are the minimum and maximum heights at which both pieces reach.
The $\mathcal{W}_{pnl}$ of the open curve then can be computed based on this equation using the \citet*{Prior2016} code available online.\footnote{\url{https://www.maths.dur.ac.uk/~ktch24/code.html}} This code can also compute the $\mathcal{W}_{pl}$ and $\mathcal{W}_{r}$, which are used to calculate the linking number ($\mathcal{L}_{k} =\mathcal{W}_{r} + \mathcal{T}_{w}$, see the inset panel of Figure~\ref{f5}(a)).

The definition of the other parameter, the twist number ($\mathcal{T}_w$), is 
defined as follow: Let a smooth curve $\mathbf{y}(s)$ wrap around the central axis $\mathbf{x}(s)$, where $s$ is the arc length starting from a reference point on this central axis. $\mathbf{T}(s)$ is the unit tangent vector to the axis curve $\mathbf{x}(s)$, and $\mathbf{V}(s)$ denotes a unit vector normal to $\mathbf{T}(s)$ and points to $\mathbf{y}$ at the point
$\mathbf{y}(s) = \mathbf{x}(s) + \epsilon \mathbf{V}(s)$. Then $\mathcal{T}_w$ density can be calculated following 
the formula \citep{Berger2006,Guo2013},

\begin{equation}
\dfrac{d\mathcal{T}_{w}}{ds} = \frac{1}{2\pi }\mathbf{\textit{T}}\cdot \mathbf{\textit{V}}\times \dfrac{\mathbf{\textit{dV}}}{ds}, \label{eqtw}
\end{equation}

The total twist is the integration of Equation~\ref{eqtw} along the axis curve $\mathbf{x}$. 
If the field line is in the vicinity of the MFR's axis and the MFR is approximately cylindrically symmetric, 
it is more convenient to use the twist number of an individual magnetic field line defined by

\begin{equation}
\mathcal{T}_{w}' = \int_{L}\frac{\mu_{0}J_{\parallel }}{4\pi B}dl=\int_{L}\frac{\curl{\mathbold{B}}\cdot\mathbold{B}}{4\pi B^2}dl \label{eqtw1}
\end{equation}
$\mathcal{T}_w'$ measures how much two neighboring close field lines twist about each other. It is a reliable approximation of the twist number
with respect to the axis, $\mathcal{T}_w$, as computed by integration of Equation~\eqref{eqtw} \citep[][Appendix C]{Liu2016}.
The total twist is the line integral of the twist intensity ($\curl{\mathbold{B}}\cdot\mathbold{B}/4\pi B^2$) along each individual field line.

\begin{figure*} 
     \vspace{-0.0\textwidth}    
     \centerline{\hspace*{0.00\textwidth}
     \includegraphics[width=1.0\textwidth,clip=]{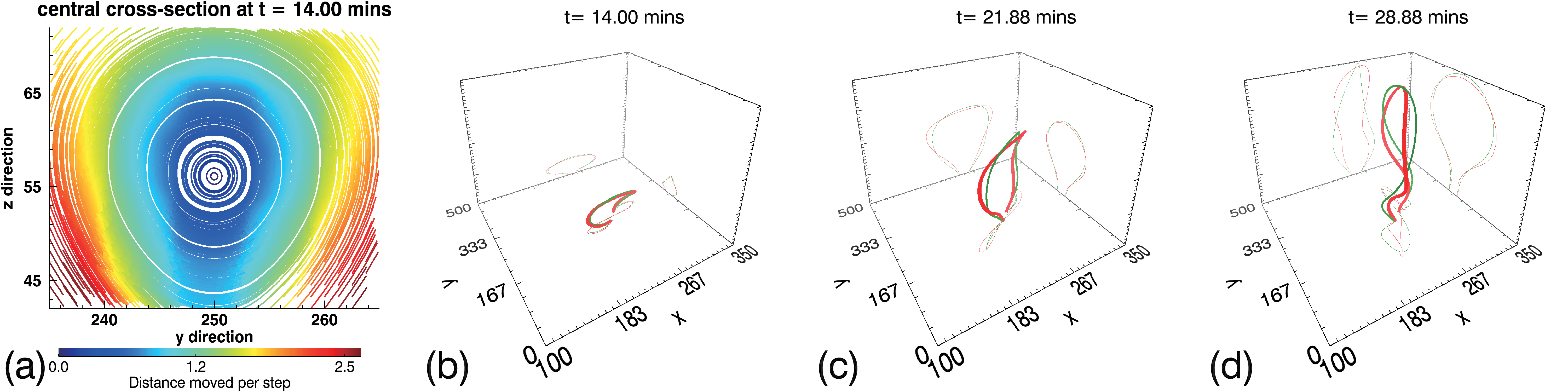}
               }
\caption{(a) Streamlines of the magnetic field at central plane at t = 28 mins, showing the cross-section of the MFR; Panels (b)-(d) Evolutions of the MFR's axis (red) and one twisted field line (green) wrapping around.} \label{f4}
\end{figure*}

To calculate the two parameters, the top priority is the determination of the MFR's axis. 
Due to the symmetry of modeled MFR's geometry, the streamlines of its transverse magnetic field forms a series of concentric rings at the central cross-section (Figure~\ref{f4}(a)), and the axis of the MFR can be identified clearly as the field line passing through the center of these concentric rings.
Furthermore, the bottom surface of the MHD model is fixed without any motion during the eruption, thus for any field line without reconnection, its two footpoints will not change with time owing to this line-tied boundary condition. Therefore, once the axis is located initially, its subsequent evolution can be followed by tracing the field line from one fixed footpoint of the axis (red line in Figure~\ref{f4}(b)-(d)). To compute the twist number, we traced eight sample field lines around the axis starting from eight points in the neighborhood surrounding the footpoint of the axis (Figure~\ref{f4}(b)-(d), as an example for one of the eight field lines). During the eruption, the other footpoints' location of the axis and wrapping field line moves less than 1.8 grid points, lower than 5.8\% changing rate relative to the separation of their two footponts.

Accompanied with the MFR rising and rotation motions (Figure~\ref{f2}(a) and (c)), temporal evolutions of $\mathcal{W}_{pnl}$ and $\mathcal{T}_{w}$ of the neighboring field line are shown in Figure~\ref{f5}. Initially, the reverse S-shaped MFR possesses a positive $\mathcal{W}_{pnl}$ of $0.33$, 
and as the CCW rotation of the apex of the MFR about its rise direction sets in, 
this value decreases monotonically to $0$, indicating that the initial strong reverse S-shaped bending is completely straightened out. Moreover, as the rotation goes on, $\mathcal{W}_{pnl}$ even reverses its sign to a negative value of $-0.07$. On the other hand, the neighboring field lines winding around the axis have initially a left-handed twist on average of ($\mathcal{T}_{w}$) $\approx - 1.70$, and as the eruption goes on, the twist is enhanced with its absolute value growing gradually to $2.09$, which indicates that the CCW rotation motion twists up these spiral field lines. 
Based on the extension for the open field lines of C\u{a}lug\u{a}reanu theorem \citep{Berger2006}, the total linking number ($\mathcal{L}_{k} = \mathcal{T}_{w} + \mathcal{W}_{r}$) is proved mathematically invariant to all motions (as long as no reconnection happens between these field lines). This is consistent with our result (the changing rate of $\mathcal{L}_{k}$ is less than 8\%, see the inset panel of Figure~\ref{f5}(a)) which shows a negative correlation between $\mathcal{W}_r$ and $\mathcal{T}_{w}$ and indicates that writhe transfers to twist during the eruption. 
We have also checked the energy and helicity evolution during the simulated eruption. As can be seen in the bottom panel of Figure~\ref{f5}, the free magnetic energy in the volume is rapidly released by 50\% through the eruption, while the magnetic helicity is preserved pretty well in the volume with only a small variation of less than 2\%.

\section{Discussion and Conclusion} \label{sec:dis}
In this Letter, the relationship between the direction of filament rotation during eruption, the orientation of S-shaped morphology of filament (and co-spatial sigmoid), and the chirality of the filament is studied using a fully 3D MHD simulation of solar eruption initiation. 
The simulated flux rope eruption resembles the initial morphology and rotation motion of an erupted filament. The emission image as synthesized from the electric current density in the model shows an inverse sigmoidal pattern in the wake of the eruption and the sigmoid-to-arcade transformation. Furtherly, $\mathcal{W}_{pnl}$ and $\mathcal{T}_{w}$ of the simulated MFR, the quantitative parameters describing the deformation of the axis and its wrapping field lines, are calculated, which shows clearly an accumulation of the $\mathcal{T}_{w}$ and a reduction of the $\mathcal{W}_{pnl}$ during the eruption. Such a transfer from writhe to twist is at variance with the existing explanation for MFR rotation invoking the helical kink instability in which the twist of MFR is converted to writhe.

Many attempts have been made before to
determine this observed rotation--chirality relationship. For example,
\citet{Green2007} has comprehensively 
reviewed various models of sigmoid formation and considered this observed property
as a consequence of the conversion of twist into writhe under the ideal MHD constraint of helicity conservation.
But this leaves a mystery: 
through the rotation the original inverse S-shaped filament spine is straightened and even over rotated to become forward S-shaped, which contradicts the expectation of the kink instability. Some observation and simulation suggest that eruption of a low-lying MFR with downward-bent axis also accommodates this scenario \citep[e.g.,][]{Torok2010,Zhou2017}. But for the studied filament eruption, no obvious dip (i.e., with a concave-upward motion) is present in the middle, and the initial reverse S-shape of the filament is formed largely by the two curved ends rather than downward-bent or flat portions in its mainbody.  
This mystery is solved in our analysis here, that a forward (reverse) S-shaped filament eruption shows CW (CCW) rotation is consistent with the eruption scenario as demonstrated by \citet{Jiang2021}'s simulation, namely, the erupting MFR is formed during the eruption by tether-cutting reconnection. 
The physics behind the key different behavior of MFR formed during eruption from that formed prior to eruption will be investigated in future works.

\section{acknowledgements } \label{sec:ack}
The authors wish to express their special thanks to the referee for suggestions
and comments which led to the improvement of the paper. 
The authors appreciate discussions with Guo Yang, Xin Cheng, and Xudong Sun.
We acknowledge the \emph{SECCHI}, \emph{AIA}, \emph{GONG}, \emph{XRT}, and \emph{HMI} consortia for providing  excellent observations.

This work is supported by the B-type Strategic Priority Program XDB41000000 funded by the Chinese
Academy of Sciences. The authors also acknowledge support from the National Natural Science Foundation of China (NSFC 42004142, 41822404, 41731067, 11925302, 42188101, 41822404, 41731067, 41574170, and 41531073), Open Research Program of CAS Key
Laboratory of Geospace Environment, the Fundamental Research Funds for the Central Universities (grant No. HIT.BRETIV.201901), and Shenzhen Technology Project JCYJ20190806142609035.



\end{CJK*}
\end{document}